\documentclass{article}

\usepackage{arxiv}

\usepackage[utf8]{inputenc} 
\usepackage[T1]{fontenc}    
\usepackage{hyperref}       
\usepackage{url}            
\usepackage{booktabs}       
\usepackage{amsfonts}       
\usepackage{nicefrac}       
\usepackage{microtype}      
\usepackage{lipsum}		
\usepackage{graphicx}
\usepackage{natbib}
\usepackage{doi}
\usepackage{amsmath}

\hypersetup{
    colorlinks=true,  
    linkcolor=black,  
    citecolor=black,  
    urlcolor=blue     
}

\title{Persistent Homology Broadens the Controllable Subspace in Human Structural Connectomes}

\author{
\href{https://orcid.org/0009-0008-7120-1797}{\includegraphics[scale=0.06]{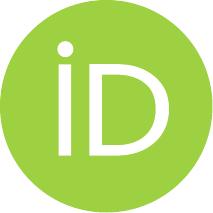}\hspace{1mm}Carter Sale}$^{1,2,3}$\thanks{Correspondence: carter.sale@mq.edu.au},
\href{https://orcid.org/0000-0000-0000-0000}{\includegraphics[scale=0.06]{orcid.pdf}\hspace{1mm}Marco Coraggio}$^{1}$,
\href{https://orcid.org/0000-0003-3085-152X}{\includegraphics[scale=0.06]{orcid.pdf}\hspace{1mm}Mengsen Zhang}$^{4}$, 
\href{https://orcid.org/0000-0001-9159-2774}{\includegraphics[scale=0.06]{orcid.pdf}\hspace{1mm}Michael J. Richardson}$^{2,3}$, \\
$^{1}$ Modeling and Engineering Risk and Complexity, Scuola Superiore Meridionale, Italy \\
$^{2}$ School of Psychological Sciences, Macquarie University, Australia \\
$^{3}$ Performance and Expertise Research Centre, Macquarie University, Australia \\
$^{4}$ Department of Computational Mathematics, Science, and Engineering, Michigan State University, USA \\
}

\date{} 

\renewcommand{\shorttitle}{}

\hypersetup{
pdftitle={},
pdfsubject={q-bio.NC, q-bio.QM},
pdfauthor={Carter Sale},
pdfkeywords={},
}

\begin{document}
\maketitle

\begin{abstract}
Network control theory applied to structural connectomes typically ranks brain regions as candidate driver nodes by their structural connectivity strength, and evaluates performance through scalar control energy. We test whether this framing captures the most relevant information about how driver-node selection shapes brain network control. We introduce an alternative criterion based on the persistent topological cycles in which each node participates---a measure of mesoscale integration that captures features beyond local connectivity---and compare it to standard degree-based selection across 70 human structural connectomes at three parcellation scales. Topology- and degree-informed driver sets achieve nearly identical scalar control energy, differing by approximately 0.2\%. The geometry of the controllable subspace, however, differs substantially: topology-informed sets distribute controllability across more dimensions of state space and produce better-conditioned controllability matrices. This geometric advantage is preserved when high-degree hub nodes are removed, and it carries a functional signature: because the two criteria place driver nodes in different cortical territory, each most efficiently reaches a different class of target state. The choice of node-ranking criterion therefore shapes which brain-state transitions are energetically favored even when average control cost is unchanged. The results reveal a dissociation between control cost and control geometry, and demonstrate that persistent topology captures information about brain network control that scalar energy summaries miss.
\end{abstract}


\section{Introduction}

The human brain can be naturally modeled as a dynamical system in which neural activity evolves over a fixed structural substrate, with white-matter connections linking distributed cortical and subcortical regions \cite{sporns2005human,bullmore2009complex}. This perspective motivates the application of network control theory to structural connectomes derived from diffusion MRI, where the connectivity matrix is interpreted as the state matrix of a linear time-invariant system and external inputs are used to steer network activity between states \cite{pasqualetti2014controllability,tang2018colloquium}. Here a central quantity of interest is the controllability Gramian, whose trace encodes the average energy required to reach any point in state space from the origin. Empirical studies have used this and related scalar summaries to show that brain regions differ in their capacity to drive the network toward certain states, and that these differences are tied to the region's structural embedding, its position in development and disease, and its relevance for cognitive function and noninvasive brain stimulation \cite{gu2015controllability,betzel2016optimally,tang2018colloquium}. Node-level rankings based on local graph metrics have become a natural principle in this literature. Average and modal controllability correlate strongly with weighted degree \cite{gu2015controllability,betzel2016optimally}, the degree distribution determines how many driver nodes a network requires for structural controllability \cite{liu2011controllability}, and degree-weighted centrality measures have been proposed as explicit heuristics for minimizing control energy in complex networks \cite{lindmark2018minimum,pasqualetti2014controllability}. Brain network control theory (NCT) has largely adopted the same logic: the standard pipeline ranks nodes by a graph-theoretic criterion, selects the top-$k$ as the input set, and evaluates performance through scalar control energy \cite{parkes2024network}. 

This node-centric, energy-focused framing has proven productive, but it contains two conceptual gaps that motivate the present work. First, the controllability Gramian is a matrix, and collapsing it to a scalar---typically its trace or the trace of its inverse---discards information about the geometry of the controllable subspace \cite{pasqualetti2014controllability,liu2011controllability,liu2016control}. The trace of $W^{-1}$ reflects average control cost, but the eigenspectrum of $W$ describes which directions in state space are easy or difficult to reach and how much of that space is accessible at all \cite{kim2018role}. Two driver sets that produce identical scalar energy can differ substantially in whether they provide access to a broad or narrow region of state space. Second, the input matrix $B$---which specifies which nodes receive external input---is a set-level object: controllability depends on the joint configuration of driver nodes, not on the sole contribution of any individual node. Standard rankings, which evaluate nodes in isolation, are potentially ignoring synergies or redundancies among them. This matters because the complementarity of a driver set---the degree to which adding new nodes extends coverage of state space rather than overlapping with directions already controlled---is not recoverable from node-level scores alone, and need not track with those scores at all \cite{kim2018role}. Indeed, structural controllability results show that driver nodes counterintuitively tend to avoid the highest-degree hubs \cite{liu2011controllability}, and empirical work on multi-input control of structural brain networks finds that spatially distributed sets outperform hub-clustered ones, since redundant coverage among nearby hubs leaves distant regions poorly reachable \cite{alizadeh2024impact}. These results suggest that the connection between degree and driver-node quality is not straightforward. 

Both gaps point toward missing information about the collective, mesoscale organization of the network---how nodes participate in cycles, loops, and higher-order structures that span multiple regions and cannot be identified from any single node's local connectivity. A natural tool for extracting such information is persistent homology, which tracks topological features---connected components, cycles, and higher dimensional cavities---across a filtration of the network and summarizes them in terms of their birth and death thresholds \cite{petri2014homological,sizemore2018cliques}. Applied to the clique complex of a structural connectome, $H_1$ persistent homology identifies cycles: closed paths through the network whose existence reflects mesoscale loop structure rather than local hub density \cite{giusti_cliquetopologyreveals_2015,giusti2016two}. Nodes that participate in many persistent cycles are embedded in this mesoscale architecture in a way that is structurally distinct from nodes that are simply highly connected; the two properties need not coincide \cite{battiston2020networks,sizemore2018cliques}. We therefore introduce cycle participation as a driver node selection criterion: nodes are ranked by the number of persistent $H_1$ cycles in which they appear as representatives, giving priority to nodes whose structural position bridges and anchors the network's higher-order loop structure. The key question is not whether cycle participation outperforms degree as an energy minimizer but whether it captures combinatorial and geometric information about control that degree misses.

We pursue this question in 70 individual human structural connectomes parcellated at three resolutions---68, 114, and 219 regions---using the Lausanne multi-scale atlas \cite{cammoun2012mapping}. Each connectome was modeled as a continuous-time linear time-invariant system whose state matrix was the stabilized, log-transformed, and range-normalized weighted adjacency matrix derived from deterministic streamline tractography. Controllability was characterized via the infinite-horizon controllability Gramian, evaluated both as a scalar summary (trace of the inverse, proportional to average control energy) and as a spectral object (effective rank, participation ratio, and condition number). Driver nodes were selected by two criteria applied to the same preprocessed adjacency matrix: degree strength, defined as the weighted sum of incident edges, and cycle participation, defined as the persistence-weighted count of $H_1$ representative cycles in which a node appears, computed from the Vietoris--Rips filtration of the clique complex. 

Our analysis proceeds in two parts. We first establish a methodological observation that motivates the geometric framing of what follows: the scalar control-energy landscape becomes increasingly degenerate with parcellation resolution---at fine scale, the distribution of average energies across candidate driver sets narrows sharply, and node identity becomes largely irrelevant to scalar energy as a selection criterion. This degeneracy motivates a shift from energy alone toward geometric characterization of the controllable subspace. We then report three findings. First, topology-informed driver sets achieve nearly identical average control energy to degree-informed sets but produce consistently broader controllable subspaces---higher effective rank, higher participation ratio, and lower Gramian condition numbers. Second, this geometric advantage is preserved under targeted removal of high-degree hub nodes: degree-informed sets lose substantial spectral breadth while topology-informed sets are largely resilient, despite equivalent energy degradation in both strategies. Third, the geometric advantage translates into target-specific reductions in transition energy: topology-informed sets lower transition energy for targets weighted toward visual cortex, whereas degree-informed sets are more efficient for association and sensorimotor targets. Together, these results suggest that higher-order topology contributes less by reducing average control cost than by organizing driver sets with broader geometric reach---an axis of network control organization that scalar energy summaries miss, and one that is robust to the hub disruption characteristic of focal injury and neurodegeneration.

\section{Results}

\subsection{Topology-informed and degree-based rankings emphasize overlapping but nonidentical cortical territories}

\begin{figure}
    \centering
    \includegraphics[width=0.8\linewidth]{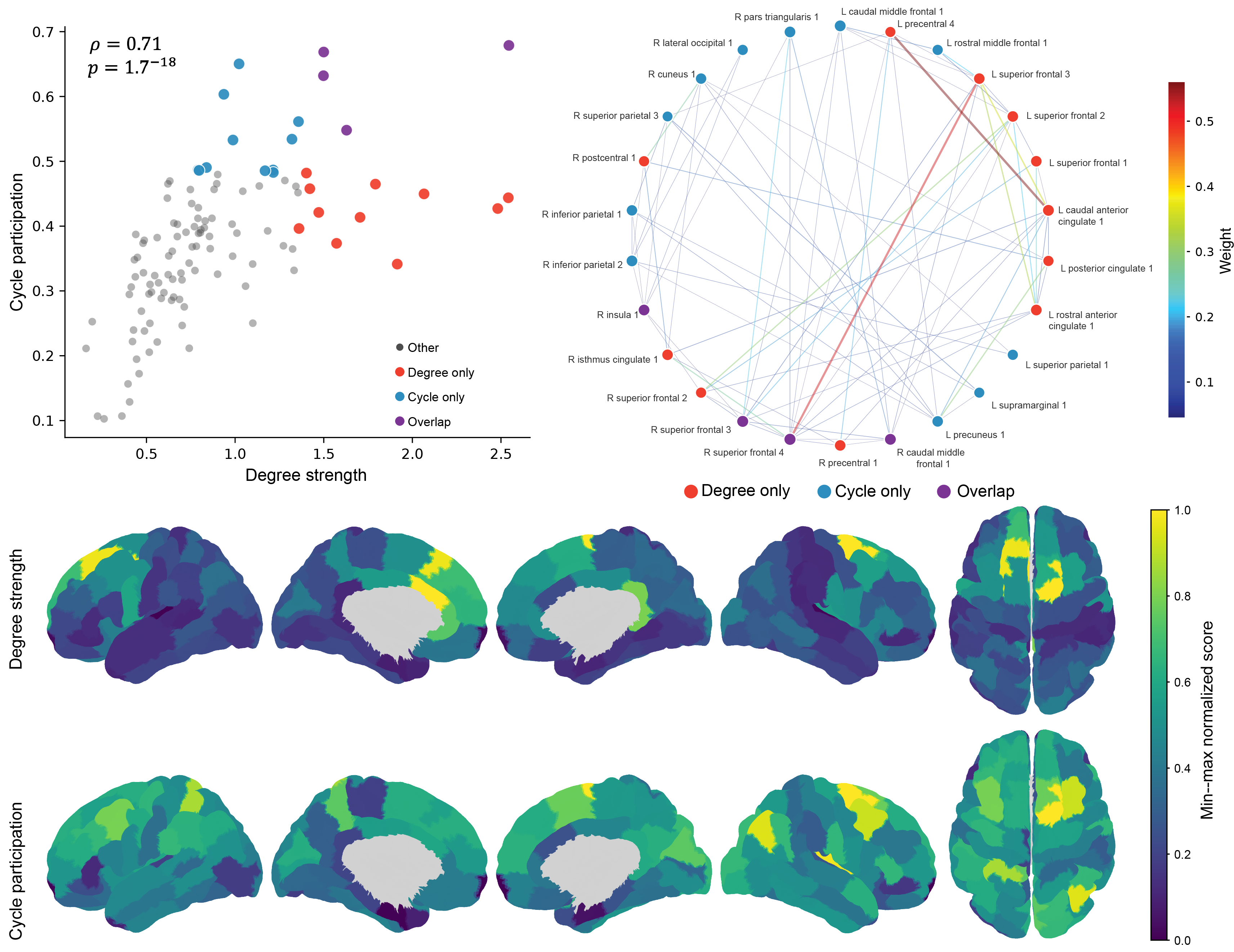}
    \caption{\textbf{Topology-based and degree-based node rankings emphasize overlapping but nonidentical territory.} All panels use subject-averaged scores at the 114-region parcellation scale. Top left: mean cycle participation versus mean degree strength across parcels. The two are strongly correlated (Spearman $\rho = 0.71$, $p = 1.7\times10^{-18}$, $n = 114$) but with substantial dispersion, so parcels with similar degree can differ markedly in cycle participation. Points are colored by membership in the top-15 degree set only, the top-15 cycle set only, both (overlap), or neither (other). Top right: connectivity subgraph induced by the union of the top-15 degree and top-15 cycle-participation parcels, arranged on a circle and labeled by region; node color denotes degree-only, cycle-only, or overlapping selection, and edges are colored and weighted by structural connection strength (colorbar). Bottom: mean degree strength (upper row) and cycle participation (lower row) rendered on the cortical surface (fsaverage surface; left-hemisphere lateral and medial, right-hemisphere medial and lateral, and dorsal views), each min--max normalized to $[0,1]$ on a shared color scale, with the medial wall in gray.}
    \label{fig:rank_divergence}
\end{figure}

We first examined the anatomical distribution of cycle participation to determine whether the topological signal identifies meaningful brain regions or merely reflects arbitrary graph structure. At the 114-region parcellation scale shown in Figure~\ref{fig:rank_divergence}, high cycle-participation values were concentrated in superior and middle frontal, inferior and superior parietal, precuneus, supramarginal, insular, and occipital (cuneus and lateral occipital) cortex. These regions span association and integrative cortical territories together with medial and lateral visual cortex, foreshadowing the visual bias of cycle participation described below.

Despite this anatomical overlap with known hub regions, cycle participation was not reducible to degree: specifically, across nodes in the subject-averaged 114-region table used for the main figure, cycle participation and degree strength were strongly positively correlated (Spearman \(\rho = 0.71\), \(p = 1.7 \times 10^{-18}\), \(n = 114\); Fig.~\ref{fig:rank_divergence}), but substantial dispersion around the trend indicated that regions with similar degree could differ markedly in their participation in persistent \(H_1\) structure. A similar positive association was present at the 68- and 219-region scales (68: $\rho=0.75, p<.001$; 219: $\rho=0.61, p<.001$). However, the coarsest scale showed substantially greater agreement in the highest-priority regions, with 11 of the top 15 cycle-participation regions also appearing among the top 15 degree-ranked regions at 68 regions, compared with only 4/15 at 114 regions and 5/15 at 219 regions. Thus, cycle participation identified anatomically meaningful regions while retaining information not fully explained by local degree, particularly at finer parcellation scales where the two rankings diverged more.

Displayed as separate cortical surfaces, the two rankings emphasized different territory (Fig.~\ref{fig:rank_divergence}). To relate this difference to macroscale cortical organization, we correlated the parcelwise rank divergence (degree rank minus cycle rank) with the principal cortical gradient \cite{margulies2016situating}. The divergence was negatively correlated with the gradient at all three scales (Spearman \(\rho = -0.23\), \(-0.38\), and \(-0.32\) at 68, 114, and 219 regions): regions toward the transmodal, association end of the gradient were prioritized by degree strength, whereas regions toward the unimodal, visual end were prioritized by cycle participation. This anatomical dissociation anticipates the target-state results reported below, in which each criterion most efficiently reaches target states aligned with the territory it prioritizes.

\subsection{Divergence between topology-based and degree-based rankings predicts relative control energy}

\begin{figure}
    \centering
    \includegraphics[width=\linewidth]{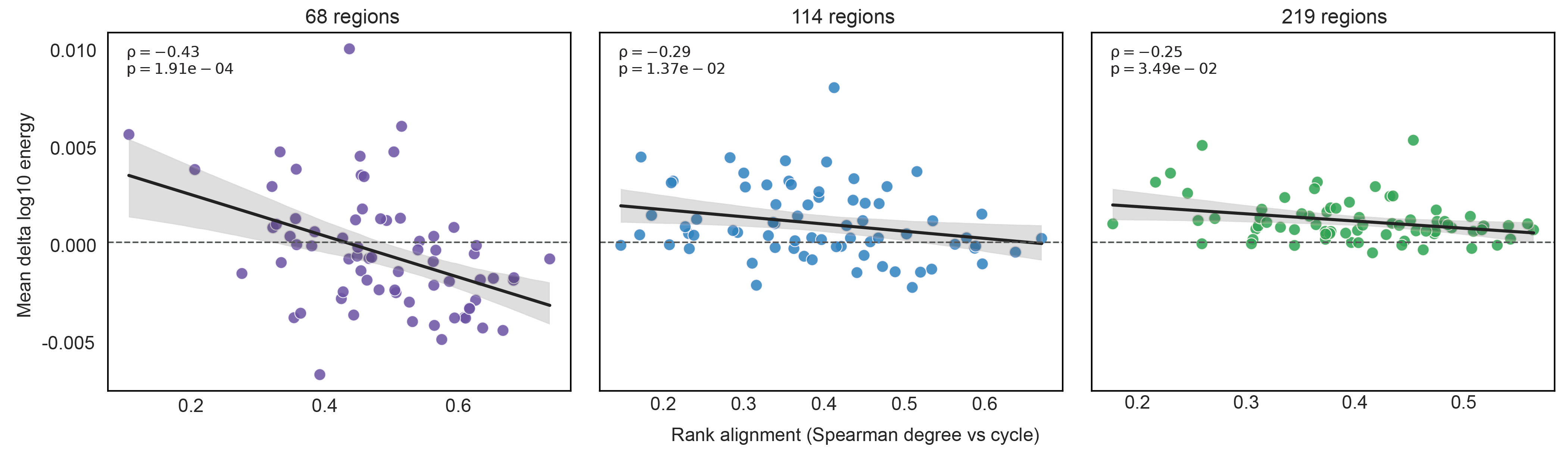}
    \caption{\textbf{Rank alignment predicts the relative energy of static topology-informed and degree-informed driver sets.} Each point represents one subject; rank alignment is the within-subject Spearman correlation between degree-strength and cycle-participation rankings, and relative energy is averaged across $k\in\{1,2,3,5\}$. Positive values of $\Delta E=\log_{10}(E_{\mathrm{topology}})-\log_{10}(E_{\mathrm{degree}})$ indicate higher average control energy for topology-informed selection. Lines show the across-subject association at each parcellation scale. Annotations show unadjusted Spearman tests; associations survived Holm correction at 68 and 114 regions, but not at 219 regions.}
    \label{fig:rank_energy_divergence}
\end{figure}

To test whether disagreement between topology-based and degree-based rankings carried consequences for control, we quantified within-subject rank alignment as the Spearman correlation between cycle participation and degree strength, and related this value to the relative average energy of the corresponding static top-$k$ driver sets. For each subject, relative energy was summarized across $k\in\{1,2,3,5\}$ as $\Delta E=\log_{10}(E_{\mathrm{topology}})-\log_{10}(E_{\mathrm{degree}})$, such that positive values indicate a disadvantage for topology-based selection.

The rankings were positively but incompletely aligned in individual connectomes, with mean within-subject rank correlations of $0.481\pm0.014$ standard error (SEM) at 68 regions, $0.386\pm0.015$ at 114 regions, and $0.395\pm0.011$ at 219 regions. Averaged across driver-set sizes, topology- and degree-informed sets had almost indistinguishable scalar energy at 68 regions ($\Delta E=-5.76\times10^{-4}$ log units; Holm-adjusted $p=0.068$). At 114 and 219 regions, topology-informed static sets required slightly higher average energy than degree-informed sets ($\Delta E=9.69\times10^{-4}$ and $1.10\times10^{-3}$ log units, respectively; Holm-adjusted $p=9.6\times10^{-4}$ and $6.7\times10^{-11}$). Although statistically reliable, these differences corresponded to small linear-energy changes of approximately $0.22\%$ and $0.25\%$.

At 68 regions, lower alignment between topology and degree rankings was associated with a larger topology-related energy disadvantage ($\rho=-0.43$, Holm-adjusted $p=9.6\times10^{-4}$). This relationship remained detectable, though weaker, at 114 regions ($\rho=-0.29$, Holm-adjusted $p=0.041$). At 219 regions, the nominal association was smaller and did not survive correction ($\rho=-0.25$, Holm-adjusted $p=0.068$). Thus, where node identity remained consequential, selecting nodes that departed more strongly from degree could modestly alter scalar control cost (Fig.~\ref{fig:rank_energy_divergence}). The attenuation of this relationship at the finest scale motivated an analysis of whether the control-energy landscape itself becomes less selective with increasing parcellation resolution.

\subsection{The control landscape becomes increasingly degenerate with parcellation scale}

\begin{figure}
    \centering
    \includegraphics[width=0.8\linewidth]{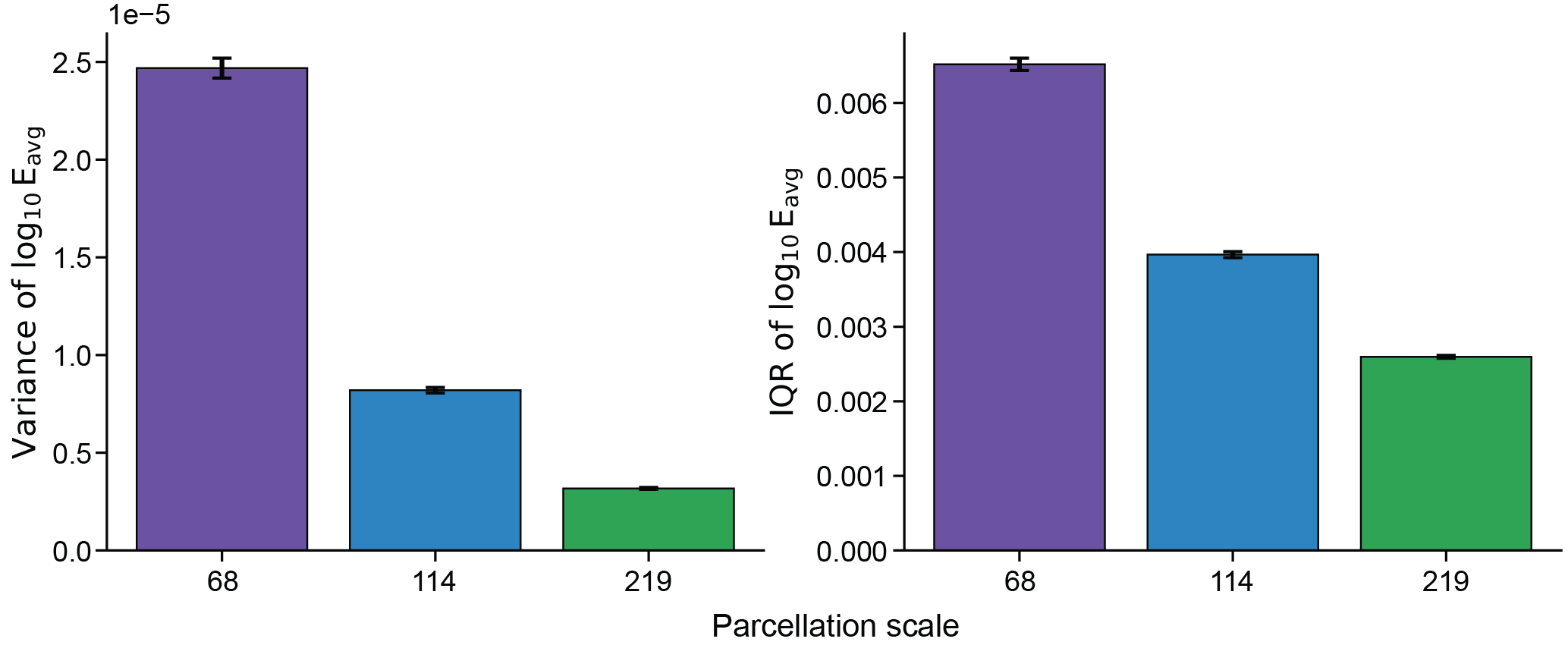}
    \caption{\textbf{The control-energy landscape becomes increasingly compressed at finer parcellation scales.} Driver sets were evaluated at fixed absolute sizes ($k\in\{1,2,3,5\}$), with all sets enumerated when feasible and otherwise 5,000 sets sampled per subject, scale, and driver-set size. Left: mean across-subject variance of $\log_{10}$ average control energy across candidate driver sets; error bars denote standard error (SEM). Right: mean interquartile range (IQR) of $\log_{10}$ average control energy across candidate driver sets. As resolution increased, both between-set variance and the central spread of the sampled energy distribution decreased.}
    \label{fig:landscape_degeneracy}
\end{figure}

The weakening relationship between rank divergence and relative energy at finer parcellation scales suggested that scalar control energy may become less sensitive to the identities of the selected driver nodes. To characterize this sensitivity directly, we evaluated the average control energy across candidate driver sets at fixed absolute sizes, $k\in\{1,2,3,5\}$. For each subject, scale, and $k$, all sets were enumerated when feasible and otherwise 5,000 unique sets were sampled.

The sampled distribution of control energies compressed substantially with increasing parcellation resolution (Fig.~\ref{fig:landscape_degeneracy}). Averaged over set sizes, the mean variance of $\log_{10}$ average energy declined from $2.47\times10^{-5}$ at 68 regions to $8.20\times10^{-6}$ at 114 regions and $3.17\times10^{-6}$ at 219 regions. The mean interquartile range showed the same monotonic compression, decreasing from $6.52\times10^{-3}$ log units at 68 regions to $3.97\times10^{-3}$ at 114 regions and $2.60\times10^{-3}$ at 219 regions. The same monotonic compression was present when the analysis was restricted to $k=1$, where the full combinatorial space was enumerated exhaustively at every parcellation scale (mean variance: $1.85 \times 10^{-5}$, $6.66 \times 10^{-6}$, and $2.47 \times 10^{-6}$; mean IQR: $5.31 \times 10^{-3}$, $3.87 \times 10^{-3}$, and $2.60 \times 10^{-3}$ log units at 68, 114, and 219 regions, respectively), indicating that the trend does not depend on the finite sample of driver sets evaluated at larger $k$. Thus, finer parcellations yielded a substantially narrower scalar-energy landscape in which average energy distinguished candidate driver sets only weakly.

\subsection{Topology-informed sets broaden control geometry}

\begin{figure}
    \centering
    \includegraphics[width=\linewidth]{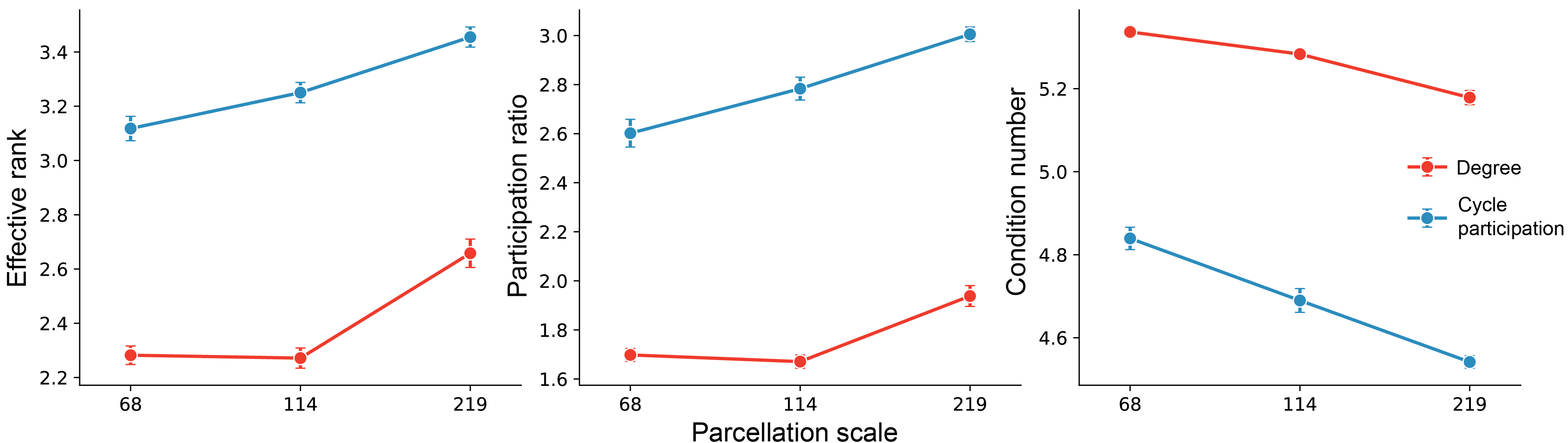}
    \caption{\textbf{Topology-informed driver sets broaden control geometry relative to degree-informed sets.} For each metric, lines show the across-subject mean for cycle-participation (blue) and degree-strength (red) static driver sets, averaged across $k\in\{1,2,3,5\}$, at each parcellation scale; error bars denote standard error (SEM). Cycle-participation sets show higher effective rank (left) and participation ratio (middle), indicating that they distribute controllability across a broader set of Gramian eigen-directions, and a lower $\log_{10}$ condition number (right), indicating a less anisotropic, better-conditioned controllability Gramian.}
    \label{fig:control_geometry}
\end{figure}

Scalar control energy summarizes controllability as a single number but discards information about which directions in state space are accessible and how evenly controllability is distributed across them. We therefore characterized the eigenspectrum of the controllability Gramian using three complementary measures: effective rank, which quantifies how broadly controllability is distributed across eigen-directions; participation ratio, a related spectral-spread measure that is larger when controllability is less concentrated in a few dominant directions; and condition number, which quantifies the anisotropy between the easiest and hardest directions to control. Higher effective rank and participation ratio, and lower condition number, each indicate that a driver set provides broader, more balanced reach across state space. 

Although topology-informed and degree-informed sets were weakly separated by scalar average energy, their control geometry differed consistently across all three parcellation scales (Fig.~\ref{fig:control_geometry}). Cycle-participation sets produced higher effective rank than degree-strength sets at every scale, with mean cycle-minus-degree differences of $0.84$, $0.98$, and $0.80$ at 68, 114, and 219 regions respectively. The median paired differences were $0.83$ (95\% CI $0.71$--$0.93$), $0.97$ ($0.84$--$1.10$), and $0.82$ ($0.68$--$0.98$). All subject-level differences were positive at each scale after averaging across $k$ (unadjusted Wilcoxon $p=3.56\times10^{-13}$; Holm-adjusted $p=3.20\times10^{-12}$). Notably, this advantage persisted at 219 regions, where candidate sets were nearly indistinguishable by scalar energy, indicating that Gramian geometry captures a dimension of driver-set quality that average energy does not.

Complementary spectral measures showed the same pattern. Cycle-participation sets had higher participation ratios at all three scales (mean differences: $0.90$, $1.11$, and $1.07$), indicating less concentration of controllability in a small number of dominant eigen-directions. Gramian condition numbers were correspondingly lower (mean $\log_{10}$ differences: $-0.50$, $-0.59$, and $-0.64$), reflecting better-conditioned and less anisotropic controllable subspaces. Topology-based selection therefore identified driver sets that supported broader geometric reach rather than lower energetic cost.

\subsection{Topology-informed control geometry is resilient to hub lesioning}

\begin{figure}[ht!]
    \centering
    \includegraphics[width=0.9\linewidth]{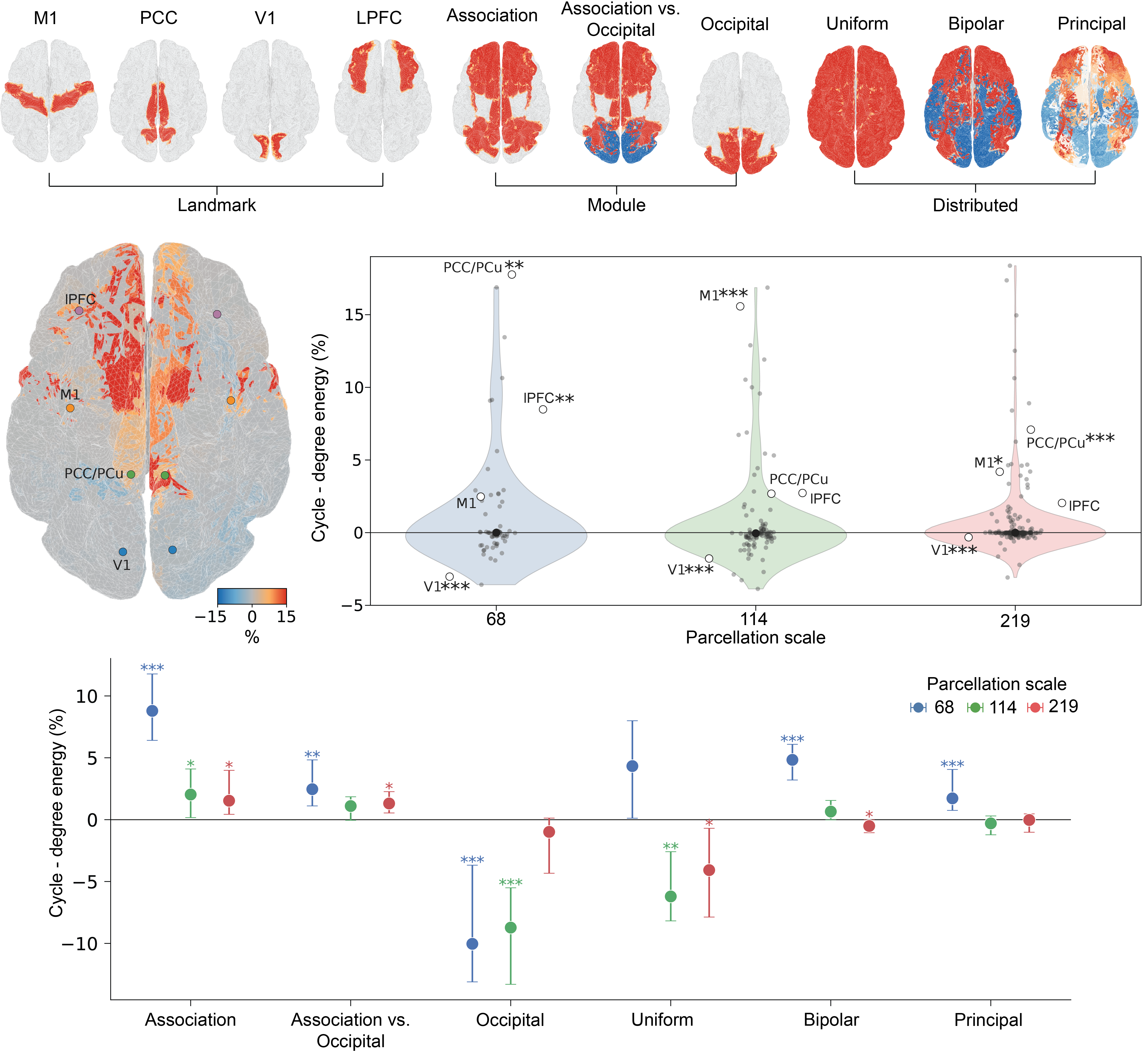}
    \caption{\textbf{Target-state control-energy differences between cycle- and degree-informed driver sets.} Top row shows the spatial definition of each target class at the 114-region parcellation scale: bilateral landmark targets, anatomical module targets, and distributed cortical activation patterns. Middle left shows the cortical single-parcel baseline, with parcels colored by subject-median percent energy difference between cycle-participation and degree-strength driver sets, ($E_{\text{cycle}} - E_{\text{degree}}$) / $E_{\text{degree}} \times 100$; blue indicates lower energy for cycle drivers and red indicates lower energy for degree drivers. Landmark targets are overlaid as labeled points. Middle right shows the distribution of cortical single-parcel effects across parcellation scales, with landmark targets overlaid as labeled points. Bottom panel shows median percent energy differences and bootstrap confidence intervals for module and distributed targets across scales. Asterisks indicate corrected significance ($*p < 0.05, **p < 0.01, ***p < 0.001$). Negative values indicate a cycle-driver advantage; positive values indicate a degree-driver advantage.}
    \label{fig:target_state_energy}
\end{figure}

Since hub nodes are disproportionately disrupted across neurological and psychiatric conditions \cite{crossley2014hubs}, we next asked whether the broader geometry of topology-informed sets depended on the integrity of the network's strongest degree hubs---and by extension, whether cycle participation identifies a structurally distinct axis of driver-node quality or merely recovers hub structure by another name. For each subject and scale, we removed the top 5, 10, or 15 degree-strength nodes and re-evaluated static cycle-participation and degree-strength driver sets selected from the remaining nodes, again across $k\in\{1,2,3,5\}$.

Hub removal disrupted degree-informed geometry substantially more than topology-informed geometry. Averaged across scales and driver-set sizes, degree-informed sets lost $0.24$, $0.32$, and $0.25$ units of effective rank after removing 5, 10, and 15 hubs respectively. Topology-informed sets showed negligible losses or modest gains ($-0.05$, $-0.06$, and $-0.11$, where negative values denote an increase). Participation ratio showed the same contrast: mean losses of $0.27$, $0.39$, and $0.34$ for degree-informed sets versus $0.09$, $0.06$, and $0.02$ for topology-informed sets. Across the 36 scale-by-lesion-by-$k$ comparisons, topology-informed sets showed significantly lower degradation after Holm correction in 32 of 36 comparisons for effective rank and 26 of 36 for participation ratio. Condition number changes were less consistent, reaching corrected significance in only 3 of 36 comparisons. The topology-related advantage was therefore expressed principally as preservation of spectral breadth under targeted hub loss.

Critically, the average energy trajectories of the two methods under hub removal were indistinguishable, replicating in a lesion context the cost--geometry dissociation observed in intact networks. Broader geometric reach may nonetheless matter only for particular directions in state space, a question we turn to next.

\subsection{Geometric differences translate into target- and scale-specific control advantages}

The average control energy $E_{\text{avg}}=\text{trace}(W_\epsilon^{-1})$ reflects the mean cost of reaching an arbitrary direction in state space, but the cost of reaching a specific target state $x^\star$ is given by $E_{\text{target}}=x^{\star \top}W_\epsilon^{-1}x^{\star}$---the projection of the inverse Gramian onto that direction. Because degree strength and cycle participation place driver nodes in different cortical territory (Fig.~\ref{fig:rank_divergence}), we expected the energetic cost of a given transition to depend on how well the target state aligns with the directions each driver set makes most controllable. 

To test this prediction, we evaluated minimum-energy transitions to a set of target states arranged along a concentrated--distributed axis. At the concentrated end, we used four bilateral anatomical landmarks chosen to span the unimodal--transmodal hierarchy---primary visual cortex (V1, pericalcarine) and primary motor cortex (M1, precentral) as canonical unimodal sensory and motor regions, and the posterior cingulate/precuneus (a core default-mode hub) and lateral prefrontal cortex (a frontoparietal-control region) as transmodal association hubs---together with an exhaustive single-cortical-parcel baseline in which each cortical parcel was evaluated as a single-node target. At the distributed end, we used three independently defined targets: the principal cortical gradient \cite{margulies2016situating}, its bipolar (sign-of-gradient) variant, and a uniform cortical activation. Intermediate targets were two anatomically defined modules (association cortex, occipital cortex) and an association--occipital contrast. All targets were defined independently of both rankings. To enable comparison across targets and scales, energy differences are reported as $100\times(E_{\mathrm{cycle}}-E_{\mathrm{degree}})/E_{\mathrm{degree}}$, with negative values indicating a cycle-driver advantage.

At the concentrated end of the axis, single-cortical-parcel targets showed effects close to zero with substantial parcel-to-parcel heterogeneity (Fig.~\ref{fig:target_state_energy}). Median percent differences across parcels were $-0.30\%$ at 68 regions, $-0.90\%$ at 114 regions, and $-0.43\%$ at 219 regions; at the two finer scales, roughly half of cortical parcels showed a significant cycle advantage and a quarter showed a significant degree advantage ($50.0\%$ and $50.2\%$ negative, $25.4\%$ and $22.8\%$ positive at 114 and 219 regions respectively). The four named landmark targets fell within the bulk of this distribution rather than at its extremes. V1 was the only landmark to consistently favor cycle drivers across scales (medians $-3.0\%$, $-1.8\%$, and $-0.3\%$ at 68, 114, and 219 regions; all Holm-adjusted $p<0.001$), while PCC/precuneus and M1 favored degree drivers at one or more scales (PCC: $+17.8\%$ at 68 regions, $+7.1\%$ at 219 regions; M1: $+15.6\%$ at 114 regions; all Holm-adjusted $p<0.001$). The lPFC landmark showed weaker and less consistent effects. This split among the concentrated landmarks---a cycle advantage for the visual target and degree advantages for M1 and the transmodal hubs---indicates that the advantage of each selection strategy tracks the target's cortical territory.

The module and whole-cortex targets confirmed that this alignment governs the advantage. The occipital module produced the largest cycle advantages of any target, at 68 and 114 regions ($-10.0\%$ and $-8.7\%$; Holm-adjusted $p<10^{-4}$), attenuating at 219 regions ($-1.0\%$). Conversely, the association module favored degree drivers at all three scales ($+8.8\%$, $+2.0\%$, and $+1.5\%$), as did the association-versus-occipital contrast, with both effects most pronounced at 68 regions. The uniform-cortical target, which weights all cortex equally, favored cycle drivers at the two finer scales ($-6.2\%$ and $-4.1\%$; Holm-adjusted $p<0.02$) but not at 68 regions. In contrast, the two gradient targets, whose activation mixes unimodal and transmodal cortex, showed no consistent advantage for either strategy: the principal-gradient target was statistically indistinguishable between strategies at the two finer scales (medians $-0.3\%$ and $-0.0\%$; Holm-adjusted $p>0.12$) and marginally favored degree at 68 regions ($+1.7\%$), and the bipolar-gradient target was similarly inconsistent in sign across scales ($+4.8\%$, $+0.7\%$, and $-0.5\%$). The topology advantage therefore tracked whether the target's activation fell on visual cortex.

Taken together, these results organize by cortical territory. Cycle-informed driver sets reduced transition energy for targets weighted toward visual cortex---V1 and the occipital module---by up to roughly 10\%, whereas degree-informed sets were more efficient for association and sensorimotor targets. This mirrors the placement of driver nodes themselves: cycle participation preferentially selects occipital and parietal regions and degree strength preferentially selects transmodal hubs, so each criterion most efficiently reaches the target states aligned with its own drivers. The uniform-cortical target, which weights all cortex equally, also favored cycle drivers at the two finer scales, consistent with their broader Gramian geometry, while the gradient targets, which load on territory favored by each criterion, showed no net advantage.

\section{Discussion}

The primary finding of this paper is that topology-informed and degree-informed driver sets differ only marginally in scalar control energy yet differ substantially and consistently in the geometry of the controllable subspace, with cycle-participation sets accessing a broader and better-conditioned region of state space across all three parcellation scales. Degree-based node rankings have become a common heuristic for driver selection in structural connectome NCT, supported by the intuition that highly connected nodes are well-positioned to propagate input throughout the network \cite{gu2015controllability,van2011rich}. Our results do not overturn this intuition---cycle-participation and degree-informed sets differed by approximately 0.2\% on average energy across the three parcellation scales we examined---but this near-equivalence in cost conceals a consistent geometric difference. Topology-informed driver sets distributed controllability across more Gramian eigen-directions, produced better-conditioned controllable subspaces, and preserved that geometric breadth under targeted hub removal, all while differing only negligibly in scalar energy. These findings suggest that the standard NCT pipeline, by collapsing the Gramian to a scalar summary, systematically discards information about what regions of state space are accessible from a given set of driver nodes \cite{kim2018role,pasqualetti2014controllability,liu2016control}.

A secondary contribution of this paper concerns the sensitivity of the energy criterion itself to parcellation resolution. At 219 regions, the distribution of energies across candidate driver sets became sharply compressed, rendering node identity essentially irrelevant to scalar energy as a selection criterion. This scale-dependent collapse persisted across stabilization constants, regularization parameters, and finite-horizon Gramians, and it was equally apparent for both proportional and fixed driver-set sizes (see Supplementary Materials). The implication is that studies that evaluate node-level controllability rankings using energy-based criteria at fine parcellation scales \cite{parkes2024network,betzel2016optimally,gu2015controllability} may be operating in a regime where the energy landscape is too flat to discriminate meaningfully among candidate driver sets. In that regime, geometric criteria retain discriminative power even when scalar energy does not. The degeneracy finding therefore motivates a methodological shift from energy minimization toward geometric characterization of the controllable subspace as a primary analysis objective in future NCT work.

The mechanistic basis for the geometric advantage of cycle-participation driver sets lies, we propose, in the structural position that high-participation nodes occupy within the network. Persistent $H_1$ cycles identify closed paths through the connectome whose existence reflects mesoscale loop structure rather than local hub density \cite{giusti_cliquetopologyreveals_2015,giusti2016two}. Nodes that appear as representatives in many persistent cycles are embedded at the intersections of multiple such loops, meaning that perturbations at those nodes propagate through structurally diverse pathways that span many eigen-directions of the system. By contrast, high-degree nodes, while strongly connected, tend to share overlapping neighborhoods that contribute correlated directions to the Gramian, concentrating controllability in a small number of dominant modes \cite{kim2018role}. This structural argument predicts the pattern observed: topology-informed sets produce higher effective rank and participation ratio not because they reduce the total energy required to reach state space but because they access more of it. Establishing this formally would require connecting the filtration structure of the Vietoris--Rips complex to the eigenspectrum of the Lyapunov Gramian, which we leave as a target for future theoretical work.

The hub-lesion analysis provides independent evidence for the cost--geometry dissociation and connects it to a biologically motivated scenario. Hub nodes in human structural connectomes---the rich-club regions identified by van den Heuvel and Sporns \cite{van2011rich,van2013network}---are disproportionately implicated across a range of neurological and psychiatric conditions. Meta-analytic evidence shows that lesion locations across 26 brain disorders preferentially overlap with structural and functional hubs \cite{crossley2014hubs}, and focal lesion studies demonstrate that hub disruption produces widespread network disconnection and multi-domain behavioral impairment \cite{siegel2018re,gratton2012focal}. In neurodegenerative disease, tau pathology and amyloid deposition spread preferentially through hub-connected cortical territories, linking connectome hub structure to the progression of Alzheimer's disease \cite{buckner2009cortical,cope2018tau}. 

The hub-removal simulation used here is an idealization of these processes, since real lesions are spatially constrained, primarily affect white matter rather than gray matter nodes, and produce network reorganization rather than simple deletion \cite{fornito2015connectomics}. Nevertheless, the simulation captures the first-order consequence of hub disruption---the loss of highly connected nodes from the controllable network---and asks how different driver-selection strategies respond to it. The answer is that degree-informed sets lose substantial geometric reach while topology-informed sets are largely preserved, despite both strategies degrading equivalently on scalar average energy. This dissociation under lesion demonstrates that the geometric advantage of topology-informed selection is robust to the removal of the very nodes that degree-based rankings prioritize, suggesting that cycle participation identifies a genuinely different structural axis of driver-node quality.

The target-state results clarify when that geometric difference matters functionally. The advantage organized by cortical territory: cycle-informed sets reduced transition energy for targets weighted toward visual cortex---V1 and the occipital module---while degree-informed sets were more efficient for association targets and for M1. The gradient targets, which load on both visual and association cortex, showed no consistent advantage for either strategy, whereas uniform cortical activation, which weights all cortex equally, favored cycle drivers at the two finer scales, consistent with their broader and less anisotropic Gramian geometry. Across all cortical parcels evaluated as single-node targets, the median energy difference was near zero at every scale, with the named landmarks (V1, M1, posterior cingulate/precuneus, lateral prefrontal cortex) falling within the bulk of this distribution rather than at its extremes.

The V1 result has a structural basis. Visual cortex is dominated by hierarchically organized feedforward and feedback projections, and the present analysis represents these as a symmetrized structural matrix, which retains their reciprocity while discarding the directional weighting. Early visual cortex therefore enters the clique complex as a densely reciprocated region that closes many short loops---exactly the structure cycle participation registers and degree strength does not. Distinguishing the contribution of symmetrization from that of genuinely recurrent visual circuitry would require a directed connectivity estimate.

That the gradient targets showed no net advantage for either strategy is itself consistent with the alignment account. The principal cortical gradient \cite{margulies2016situating} loads on occipital and sensorimotor cortex at one pole and on association cortex at the other. Only the occipital component favors cycle participation---sensorimotor and association targets both favor degree---so a transition along the gradient combines contributions that work against each other and the two strategies come out even. The one distributed target that is not tied to either end of the gradient---uniform cortical activation---did favor cycle drivers at the finer scales, which fits the broader, less anisotropic Gramian geometry of topology-informed sets: when controllability is concentrated in a small number of eigen-directions, activating many orthogonal directions is costly regardless of the total Gramian trace, whereas a broader controllable subspace spreads that cost. 

More generally, the choice of driver-selection criterion has consequences not just for average energetic efficiency but for which brain-state transitions are feasible at a given energy budget, with the advantage tracking the anatomical alignment between driver placement and target location rather than any single anatomical system. A direct empirical test---asking whether topology-informed driver regions predict individual differences in the capacity for visual state transitions, using HCP task-fMRI data or resting-state manifold geometry---represents a natural extension of the present work and a necessary step before the functional claims made here can be evaluated in vivo.

Several limitations qualify these conclusions. First, the linear time-invariant model is a strong assumption, and the debate about whether NCT metrics reflect genuine controllability of neural dynamics or are better understood as summaries of static network architecture remains unresolved \cite{tu2018warnings,suweis2019brain,pasqualetti2019re}. Although recent evidence supports the linearity of macroscopic brain dynamics at the timescales relevant to NCT \cite{nozari2024macroscopic}, the mapping between structural connectivity and effective neural dynamics is unlikely to be captured fully by a normalized adjacency matrix acting as a state-transition operator. Second, structural connectivity matrices derived from diffusion tractography carry well-characterized reliability limitations: false-positive streamlines, distance-dependent biases, and inter-algorithm variability in reconstructed connectivity \cite{maier2017challenge}. Diffusion tractography also cannot recover the directionality of white-matter pathways, so the connectome is necessarily analyzed as an undirected, symmetrized graph; the true connectivity could yield control energy and geometry that differ from the symmetrized estimate used here. Both cycle participation and degree strength are computed from the same tractography-derived matrix, so errors in the structural estimate propagate into both rankings; however, the sensitivity of $H_1$ cycle representatives to small perturbations in edge weights is a specific concern for persistent-homology-based metrics that does not apply equally to degree. 

Third, the hub-removal analysis simulates targeted attack rather than the spatially constrained, white-matter-mediated, and dynamically compensated disruption that characterizes real focal lesions and neurodegeneration \cite{fornito2015connectomics}. Results from the simulation should be interpreted as establishing a proof of principle rather than as a quantitative prediction about clinical outcomes. Fourth, the cocycle representatives returned by Ripser for $H_1$ features are not canonical: different algorithmic choices can yield different representative cycles for the same topological feature, and cycle-participation scores therefore have a degree of implementation dependence that degree strength does not. The geometric effect did, however, persist across alternative representative constructions and implementations (see Supplementary Material).

The findings reported here instantiate a more general principle that higher-order topological structure in an interaction network carries information about collective dynamics that pairwise graph metrics miss. Applied to structural connectomes, $H_1$ cycle participation captures something about the mesoscale loop architecture of white-matter connectivity that degree strength, by construction, cannot---and that something turns out to be predictive of control geometry rather than control cost. The dissociation between cost and geometry identified here is not specific to the NCT framework or to the brain: it is a general property of linear systems whose input matrices are chosen according to criteria that emphasize different aspects of network organization. Whether analogous dissociations appear in other biological control problems---gene regulatory networks, metabolic networks, spinal motor circuits---and whether persistent homology is the right tool for identifying them in those contexts, are open questions. Within the brain, the next empirical step is to anchor the geometric predictions made here to measured behavior or disease state, using larger datasets with behavioral profiling and longitudinal structural connectivity to test whether the control geometry differences identified in this paper translate into differences in cognitive capacity, resilience to injury, or response to neuromodulation.

\section{Methods}

\subsection{Structural connectome dataset and preprocessing}

\begin{figure}
    \centering
    \includegraphics[width=\linewidth]{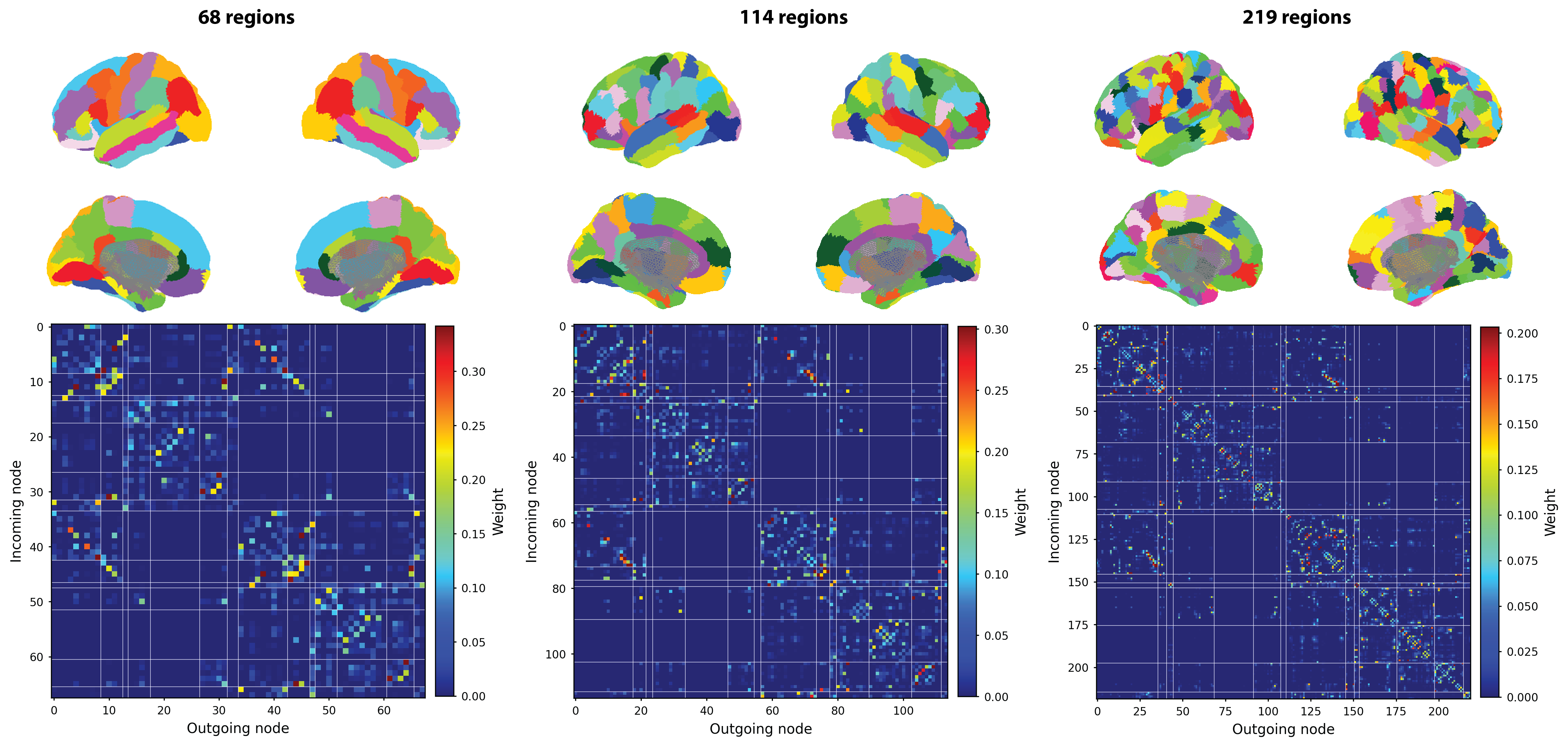}
    \caption{\textbf{Multi-scale structural connectome representations used in the analysis.} Columns show the Lausanne/Cammoun parcellations at 68, 114, and 219 regions. For each spatial resolution, cortical surface views depict the parcel boundaries for the left and right hemispheres from lateral and medial perspectives (top), and a representative weighted structural connectivity matrix is shown below (bottom). Matrix color encodes inter-regional structural connection weight, with warmer colors denoting stronger connections. Increasing parcellation resolution subdivides cortical territory into more nodes and yields a higher-dimensional structural network on which the control analyses were performed.}
    \label{fig:parcellation_scales}
\end{figure}

We analyzed diffusion MRI structural connectomes from 70 healthy adult participants \cite{griffa_2019_2872624}. Structural connectivity matrices were derived in the original dataset release using deterministic streamline tractography and normalized fiber-density estimates. We studied three spatial resolutions from the Lausanne/Cammoun multi-scale parcellation, comprising 68, 114, and 219 regions (Fig.~\ref{fig:parcellation_scales}). These three scales were chosen to test whether the inferred control architecture of the connectome depends on the granularity at which the network is represented.

For each subject and scale, we extracted a weighted structural adjacency matrix \(A_{\mathrm{raw}} \in \mathbb{R}^{N \times N}\), where \(N\) denotes the number of regions. Because the original structural matrices were undirected, all analyses were performed on symmetric weighted graphs. Each matrix was symmetrized as $A_{\mathrm{sym}} = \frac{1}{2}(A_{\mathrm{raw}} + A_{\mathrm{raw}}^\top),$ the diagonal was set to zero, and nonzero edge weights were transformed using \(\log(1+w)\) to reduce skew in the weight distribution. The resulting matrix was then normalized by its maximum entry,

$$A_0 = \frac{\log(1 + A_{\mathrm{sym}})}{\max_{ij}\log(1 + A_{\mathrm{sym},ij})},$$

yielding a weighted adjacency matrix with entries in \([0,1]\). All node-level, set-level, and control analyses were performed on these subject-specific normalized matrices.

For analyses that require a single group-representative connectome rather than per-subject matrices---the anatomical rank-divergence visualization (Section~\ref{subsec:anatomical}) and the construction of fixed target states (Section~\ref{subsec:targets})---we defined a cohort-reference adjacency matrix as the entrywise mean of the subject-specific normalized matrices,

$$\bar A_0 = \frac{1}{S}\sum_{s=1}^{S} A_0^{(s)},$$

where \(A_0^{(s)}\) is the normalized adjacency matrix of subject \(s\) and \(S=70\) is the number of subjects. The cohort reference was computed separately at each parcellation scale.

\subsection{Continuous-time linear control model}

Each connectome was modeled as a continuous-time linear time-invariant system
$$\dot{x}(t) = A x(t) + B u(t),$$

where \(x(t) \in \mathbb{R}^N\) is the network state, \(A \in \mathbb{R}^{N \times N}\) is the stabilized structural connectivity matrix, \(B \in \mathbb{R}^{N \times m}\) specifies the controlled nodes, and \(u(t)\) is the external control signal.

Because the normalized adjacency matrix \(A_0\) is not guaranteed to be asymptotically stable, we shifted its spectrum into the left half-plane by subtracting a multiple of the identity: $A = A_0 - (\lambda_{\max}(A_0) + c)I,$ where \(\lambda_{\max}(A_0)\) is the largest real part of the eigenvalues of \(A_0\), \(I\) is the identity matrix, and \(c>0\) is a fixed stabilization margin. All analyses in the main text used \(c = 0.1\). This stabilization preserves the relative network topology while ensuring that the continuous-time controllability Gramian exists.

For binary driver selection, the input matrix \(B\) was constructed by assigning one control input to each selected node, so that each column of \(B\) was a canonical basis vector corresponding to one driver region.
\subsection{Controllability Gramian and scalar control-energy metrics}

For each pair \((A,B)\), controllability was characterized using the infinite-horizon controllability Gramian \(W\), defined as the solution to the continuous Lyapunov equation $A W + W A^\top + B B^\top = 0.$

When \(A\) is stable, \(W\) summarizes how easily different directions in state space can be reached from the chosen control inputs.

Because small driver sets can yield poorly conditioned Gramians, all primary scalar and geometric summaries were computed from the regularized Gramian

$$W_\epsilon = W + \epsilon I,$$

with \(\epsilon = 10^{-5}\). After eigendecomposition, eigenvalues smaller than \(\epsilon\) were set at \(\epsilon\) for numerical stability. Our scalar control-energy proxy was

$$E_{\mathrm{avg}} = \mathrm{trace}(W_\epsilon^{-1}) = \sum_{i=1}^{N}\lambda_i^{-1},$$

where \(\{\lambda_i\}\) are the floored eigenvalues of \(W_\epsilon\). Analyses of average energy used \(\log_{10}E_{\mathrm{avg}}\).

\subsection{Node-ranking methods}

Main-text driver-selection comparisons were made between degree strength and the persistent-homology-based topology score, cycle participation.

Degree strength was defined on the preprocessed adjacency matrix as the weighted sum of incident edges, $s_i = \sum_j (A_0)_{ij},$ and nodes were ranked in descending order of \(s_i\).

Cycle participation was defined from one-dimensional persistent homology of the graph's clique complex (Section~\ref{subsec:cycle_participation}) and was likewise computed from \(A_0\). To assess whether cycle participation is redundant with classical path-based centrality, we additionally computed weighted betweenness centrality and quantified its within-subject rank correlation with cycle participation and degree strength; cycle participation was found to be no more correlated with betweenness than with degree, whereas betweenness and degree were strongly correlated with one another, indicating that cycle participation is not reducible to conventional centrality (see Supplementary Material).

For each method, the top \(k\) ranked nodes were selected as static driver sets. Unless otherwise noted, driver-set comparisons used \(k \in \{1,2,3,5\}\); the matched-proportion landscape analysis used the scale-specific set sizes described below. Scores were ordered descending using a stable sort, so tied values retained their original parcel-index order.

\subsection{Persistent-homology-based cycle participation}

\label{subsec:cycle_participation}

To compute cycle participation, each weighted adjacency matrix was converted into a distance matrix \(D\) defined by

$$    D_{ij} = \begin{cases} 1 - (A_0)_{ij}, & (A_0)_{ij} > 0, \\ 1, & (A_0)_{ij} = 0,
\end{cases}$$

with \(D_{ii}=0\). This maps strong edges to short distances while preserving absent edges as maximally distant. We computed one-dimensional persistent homology (\(H_1\)) on the resulting Vietoris--Rips filtration using Ripser with a distance-matrix input and cocycle extraction enabled \cite{bauer2021ripser}. The representative cocycle for each \(H_1\) feature is the one returned by Ripser's persistent-cohomology reduction; it is determined by filtration order of simplices, with ties broken by Ripser's combinatorial (reverse-colexicographic) indexing. The representative is therefore deterministic and reproducible for a given distance matrix but is not canonical. Geometrically, a representative cocycle is a \(1\)-cochain dual to the homological loop: its support marks edges that register winding around the corresponding one-dimensional void, and this support is not localized to the tight geodesic cycle. For this reason we define node participation from the vertices incident to the cocycle support rather than treating the (generally non-geodesic) cocycle edge set as a minimal cycle. 

For each finite \(H_1\) feature \(f\), persistence was defined as \(d_f-b_f\), where \(b_f\) and \(d_f\) are its birth and death filtration values. Node participation was defined from the endpoints appearing in the returned cocycle support \(z_f\). The unnormalized cycle-participation score was

$$\mathrm{CP}_i = \sum_{f:\, i\in\mathrm{supp}(z_f)}(d_f-b_f).$$

Scores were min--max normalized within each subject and scale before ranking. Intuitively, nodes with high cycle participation are supported by recurrent mesoscale structures that remain topologically salient across a broad range of filtration thresholds.

\subsection{Anatomical labeling and region-level summaries}
\label{subsec:anatomical}

To interpret topology-based and degree-based rankings, node indices were matched to FreeSurfer Cammoun deterministic annotation labels at each parcellation scale; parcels labeled \texttt{unknown} or \texttt{corpuscallosum} were excluded from the anatomical visualization. For Fig.~\ref{fig:rank_divergence}, degree and cycle-participation scores were first averaged across subjects at each parcel and the resulting mean scores were re-ranked. Rank divergence was defined as degree rank minus cycle-participation rank, so positive values indicate a higher priority under cycle participation. We visualized the overlap between the top 15 parcels under each score. For visualization only, the induced subgraph was drawn over union \(U\) of the two top-15 sets, displaying the strongest \(\min(n_{\mathrm{positive}},\max(35,3|U|))\) positive within-union edges, where \(n_{\mathrm{positive}}\) is the number of available positive edges. The floor of 35 and the per-node scaling of 3 (roughly three displayed edges per node) keep the diagram legible and of comparable density across parcellation scales. This choice affects only the figure and no reported quantity.  

\subsection{Node-level rank divergence and relative performance}

To test whether disagreement between topology-based and degree-based rankings predicts relative control performance, we quantified within-subject rank alignment using Spearman correlation between the node rankings induced by degree strength and cycle participation. Relative performance was defined as the difference in log control energy between the corresponding driver sets,

$$\Delta E = \log_{10} E_{\mathrm{topology}} - \log_{10} E_{\mathrm{degree}},$$

with positive values indicating a disadvantage for the topology-based set. For each subject and scale, \(\Delta E\) was averaged across static sets with \(k \in \{1,2,3,5\}\) before testing and before relating energy differences to rank alignment.

\subsection{Set-level control landscape analysis}

The primary landscape analysis used the same fixed absolute driver-set sizes as the main driver-selection comparisons, \(k\in\{1,2,3,5\}\). For each subject, parcellation scale, and \(k\), we evaluated \(E_{\mathrm{avg}}\) across candidate driver sets, enumerating all \(\binom{N}{k}\) unordered sets when that count was at most 5,000 and otherwise sampling 5,000 unique unordered sets without replacement using a fixed random seed. We summarized the resulting distribution by the variance and interquartile range of \(\log_{10} E_{\mathrm{avg}}\), which quantify how strongly scalar average energy discriminates among candidate driver sets at each scale.

As a supplementary robustness check, we also repeated the landscape analysis for driver sets of approximately equal relative size across scales, corresponding to approximately 5\% of the network: \(k=3\), \(6\), and \(11\) for the 68-, 114-, and 219-region parcellations, respectively. In these matched-proportion analyses, we sampled 5,000 unique sets per subject and scale and summarized the variance of \(\log_{10} E_{\mathrm{avg}}\).

\subsection{Target geometry and minimum-energy transitions}
\label{subsec:targets}

To examine whether different driver-selection strategies favor different classes of state transitions, we evaluated minimum-energy control to target states arranged along a concentrated--distributed axis. Targets were constructed separately at each parcellation scale from the cohort-reference adjacency matrix \(\bar A_0\) and atlas labels, then held fixed across subjects. Each target was normalized to unit Euclidean norm.

At the concentrated end of the axis, we defined four bilateral anatomical landmark targets from Cammoun base-region labels: V1/pericalcarine cortex, M1/precentral cortex, posterior cingulate/precuneus, and lateral prefrontal cortex (rostral and caudal middle frontal cortex). We also constructed an exhaustive single-parcel baseline by defining one unit-vector target at each atlas parcel; cortical single-parcel targets were used as the main concentrated baseline, with subcortical single-parcel targets retained for supplementary checks.

Intermediate anatomical targets were defined from atlas-label masks. The association mask included parcels whose labels contained \texttt{frontal}, \texttt{parietal}, \texttt{precuneus}, \texttt{cingulate}, or \texttt{insula}; the occipital mask included \texttt{cuneus}, \texttt{lateraloccipital}, \texttt{lingual}, \texttt{pericalcarine}, or \texttt{occipital}. Let \(m_S\) denote the binary indicator vector for an anatomical mask \(S\). The association-module and occipital-module targets were \(m_{\mathrm{assoc}}/\|m_{\mathrm{assoc}}\|_2\) and \(m_{\mathrm{occ}}/\|m_{\mathrm{occ}}\|_2\), respectively, and the association-versus-occipital contrast was the normalized signed vector \(m_{\mathrm{assoc}}-m_{\mathrm{occ}}\).

At the distributed end, we defined three whole-cortex targets. The principal-gradient target was the published principal functional connectivity gradient of Margulies et al. \cite{margulies2016situating}, which runs from unimodal to transmodal cortex and captures the dominant axis of macroscale cortical organization. We obtained the group-level gradient map (fcgradient01, fsLR-32k surface) through \texttt{neuromaps} \cite{markello2022neuromaps}, resampled it to the fsaverage5 surface by nearest-neighbor matching on the spherical registration, and averaged the vertexwise gradient within each Cammoun parcel to obtain a parcel-level target. Its sign was oriented so that its mean value in the association mask exceeded its mean value in the occipital mask. The bipolar-gradient target was the normalized elementwise sign of the oriented principal gradient, and the uniform-cortical target was the normalized indicator vector over all cortical parcels.

For a target state \(x^\ast\) and zero initial condition, the minimum-energy transition cost was computed as

$$E_{\mathrm{target}}(x^\ast) = {x^\ast}^{\top} W_\epsilon^{-1} x^\ast.$$

For the target-state results shown in Fig.~\ref{fig:target_state_energy}, we evaluated static degree-strength and cycle-participation sets and averaged the within-subject difference \(E_{\mathrm{cycle}}-E_{\mathrm{degree}}\) across \(k\in\{1,2,3,5\}\). Target-energy effects are reported as percent differences, \(100\times(E_{\mathrm{cycle}}-E_{\mathrm{degree}})/E_{\mathrm{degree}}\), so that negative values indicate lower transition energy for cycle-participation driver sets.

\subsection{Gramian geometry and control anisotropy}

To move beyond scalar energy summaries, we characterized the eigenspectrum of the regularized Gramian \(W_\epsilon\) for each static degree-strength and cycle-participation driver set. Let \(\{\lambda_i\}_{i=1}^N\) denote the floored eigenvalues of \(W_\epsilon\). We then computed three complementary measures of control geometry.

First, we computed the effective rank
$$r_{\mathrm{eff}} = \exp\left(-\sum_{i=1}^N p_i \log p_i\right), \qquad p_i = \frac{\lambda_i}{\sum_j \lambda_j},$$
which measures how broadly controllability is distributed across dimensions. Effective rank ranges from $1$ to $N$. It attains its minimum value of $1$ when a single eigendirection carries all controllability, and its maximum value of $N$ when all eigenvalues are equal, indicating that controllability is distributed uniformly across all state-space dimensions.

Second, we computed the participation ratio
$$\mathrm{PR} = \frac{(\sum_i \lambda_i)^2}{\sum_i \lambda_i^2},$$
which provides a related measure of spectral spread. Like effective rank, the participation ratio ranges from $1$ to $N$. It attains its minimum value of $1$ when a single eigenvalue dominates the spectrum and its maximum value of $N$ when all eigenvalues are equal, with larger values indicating that controllability is distributed more uniformly across state-space dimensions. The two measures agree at these extremes but differ in how they weight intermediate spectra: effective rank is based on spectral entropy, whereas the participation ratio is based on the ratio of the squared first and second moments of the eigenvalue distribution.

Third, we computed the Gramian condition number
$$\kappa(W_\epsilon) = \frac{\lambda_{\max}(W_\epsilon)}{\lambda_{\min}(W_\epsilon)},$$
reported on the log scale. The condition number ranges from $1$ to $\infty$ and attains its minimum value of $1$ ($\log\kappa = 0$) when all eigenvalues are equal, corresponding to a perfectly isotropic, well-conditioned Gramian in which every direction is equally controllable. It increases without bound as the spectrum becomes more anisotropic and the disparity between the most and least controllable directions grows. Thus, larger values indicate a more ill-conditioned controllable subspace, in contrast to effective rank and the participation ratio, for which larger values indicate a more uniform distribution of controllability.

For Fig.~\ref{fig:control_geometry}, metric differences were defined as cycle-participation minus degree-strength values and averaged within subject across \(k\in\{1,2,3,5\}\).

\subsection{Hub-lesion degradation analysis}

To test whether topology-informed control geometry was robust to targeted loss of conventional hubs, we performed a node-deletion analysis separately for each subject and parcellation scale. Hubs were defined from the subject-specific degree-strength ranking, and the top \(h\in\{5,10,15\}\) degree-strength nodes were removed from the preprocessed adjacency matrix. For each lesion severity and each method, we selected static driver sets of size \(k\in\{1,2,3,5\}\) from the highest-ranked nodes that survived the lesion, using the ranking computed on the intact subject-level connectome. The same surviving driver-node identities were evaluated in the intact and lesioned networks so that degradation reflected the effect of hub removal rather than reselection of a new input set.

The lesioned adjacency matrix was restabilized using the primary stabilization margin, and Gramians were computed using the same regularization procedure as in the intact analysis. For average energy and condition number, degradation was defined as the lesioned minus intact difference in the corresponding \(\log_{10}\) metric. For effective rank and participation ratio, degradation was defined as the intact minus lesioned value, such that positive values indicate loss of spectral breadth after hub removal. Strategy differences in degradation were defined as cycle-participation minus degree-strength degradation; negative values therefore indicate lower lesion-related degradation for topology-informed sets.

\subsection{Robustness analyses}

We performed supplementary robustness checks to assess whether the main energy and geometry patterns depended on numerical or sampling choices. The checks swept the following parameters, with the primary value in parentheses:

\begin{itemize}
    \item \textbf{Regularization floor}: $\epsilon \in \{10^{-8}, 10^{-7}, 10^{-6}, 10^{-5}, 10^{-4}, 10^{-3}, 10^{-2}\}$ ($\epsilon = 10^{-5}$).
    \item \textbf{Stabilization margin}: $c \in \{0.01, 1.0\}$ ($c = 0.1$), crossed with the full $\epsilon$ sweep above.
    \item \textbf{Finite-horizon Gramians}: $T \in \{1.0, 2.5, 5.0\}$, evaluated at each $c$ value with $\epsilon \in \{10^{-8}, 10^{-2}\}$ and without a regularization floor.
    \item \textbf{Larger driver-set sizes}: $k \in \{1,2,3,5,7,10,14\}$ at 68 regions, $k \in \{1,2,3,5,6,10,11,23\}$ at 114 regions, and $k \in \{1,2,3,5,10,11,22,44\}$ at 219 regions, corresponding to absolute sizes up to $k=10$ and proportional sizes of approximately 5\%, 10\%, and 20\% of the network.
\end{itemize}

We also recomputed cycle participation at 114 regions under three $H_1$ representative conventions: the Ripser cocycles used throughout \cite{bauer2021ripser}, homology pairing cycles (GUDHI birth edge and death-triangle vertices, \cite{maria2014gudhi}), and length-minimized cycles obtained by deleting the birth edge and taking the shortest replacement path at the birth filtration scale. Static driver sets and the cycle-minus-degree geometry differences were recomputed for each.

All robustness conditions used the same subject-level pairing, ranking definitions, and Gramian summaries as the primary analyses. Results are reported in the Supplementary Material.

\subsection{Aggregation and statistical analysis}

All analyses were performed at the individual-subject level and then aggregated across the full cohort of 70 subjects; the same subjects were retained at all three parcellation scales. Group summaries are reported as means \(\pm\) SEM unless otherwise stated. The main driver-set comparisons used static sets with \(k\in\{1,2,3,5\}\), and metric differences were averaged within subject and scale across \(k\) before inference.

For the scalar average-energy comparisons, two-sided Wilcoxon signed-rank tests assessed whether within-subject \(\Delta E\) differed from zero at each scale. Spearman correlation assessed the association across subjects between rank alignment and \(\Delta E\) at each scale. These six tests (three relative-energy tests and three alignment--energy associations) formed one family and were adjusted using the Holm procedure. The parcel-level correlations in Fig.~\ref{fig:rank_divergence} were descriptive comparisons of cohort-averaged scores.

For control geometry, two-sided Wilcoxon signed-rank tests assessed the cycle-minus-degree difference in effective rank, participation ratio, and \(\log_{10}\) condition number at each scale; the nine displayed tests were adjusted as one Holm family. For hub-lesion geometry, two-sided Wilcoxon signed-rank tests compared cycle- and degree-informed degradation for each of four metrics, three parcellation scales, three lesion severities, and four driver-set sizes; these 144 comparisons were Holm corrected as one family. For named target-state analyses, two-sided Wilcoxon signed-rank tests assessed \(E_{\mathrm{cycle}}-E_{\mathrm{degree}}\) and percent energy differences after averaging across \(k\); Holm correction was applied within each scale and target metric across the landmark, module, contrast, and distributed targets. For the single-parcel baseline, tests were performed parcel-wise and false-discovery-rate correction was applied within each scale, target metric, and parcel family. The landscape analysis was descriptive and did not rely on inferential tests.

\bibliographystyle{unsrt}
\bibliography{refs}  

\clearpage
\appendix
\section*{Supplementary Materials}
\setcounter{section}{0}
\renewcommand{\thesection}{S\arabic{section}}
\renewcommand{\thefigure}{S\arabic{figure}}
\renewcommand{\thetable}{S\arabic{table}}

\section{Matched-proportion landscape analysis}
\label{sec:landscape_matched}

The main text characterized landscape degeneracy using fixed absolute driver-set sizes $k\in\{1,2,3,5\}$. Because finer parcellations have more nodes, comparing the absolute width of the energy distribution across scales conflates network size with landscape shape. We therefore repeated the analysis at a fixed relative driver-set size of approximately 5\% of the network, corresponding to $k=3$, $6$, and $11$ at the 68-, 114-, and 219-region scales respectively. For each subject and scale, 5,000 unique driver sets were sampled and average control energy was computed for each. A set was defined as near-optimal if its linear energy satisfied $E_{\mathrm{avg}} \leq 1.05 \times E_{\mathrm{best}}$, where $E_{\mathrm{best}}$ is the minimum observed energy for that subject and scale.

\begin{figure}[h]
    \centering
    \includegraphics[width=0.8\linewidth]{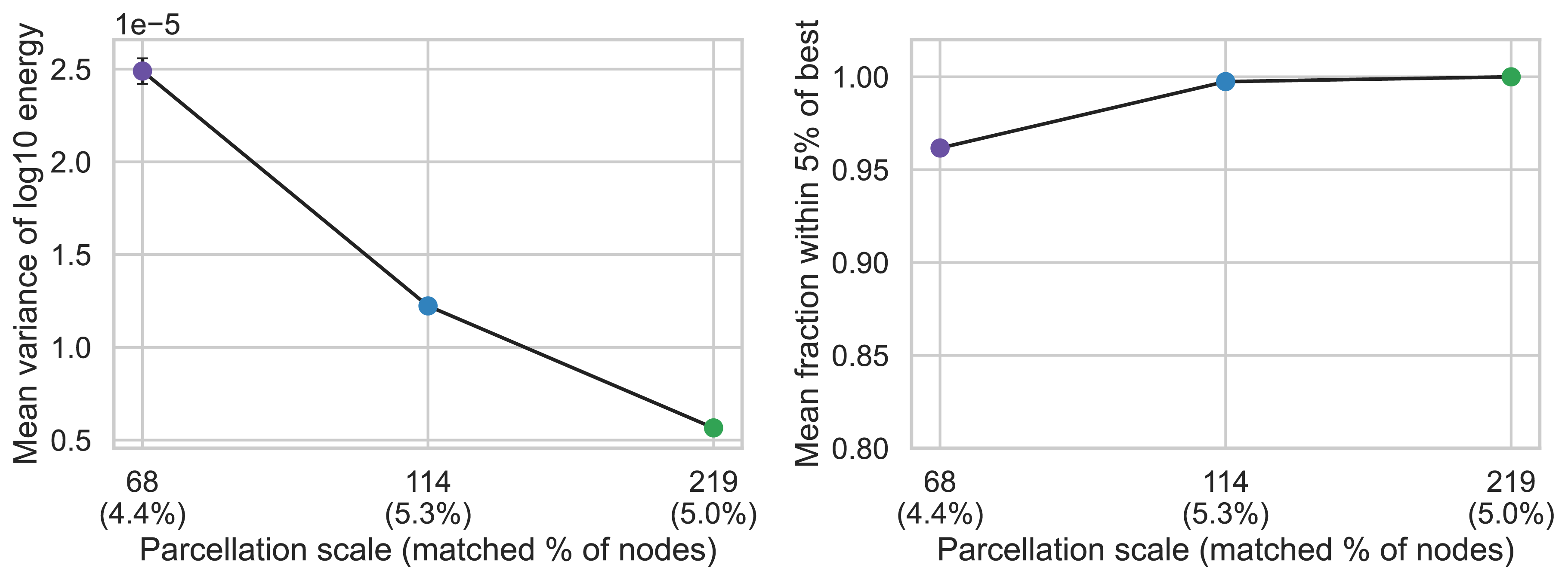}
    \caption{\textbf{Near-optimal fraction of driver sets at matched proportional size ($\approx$5\% of network).} For each subject and parcellation scale, 5,000 driver sets of size $k=3$ (68 regions), $k=6$ (114 regions), and $k=11$ (219 regions) were sampled and evaluated. Bars show the mean fraction of sampled sets within 5\% of the best observed energy; error bars denote SEM across subjects. The landscape collapses near-completely at finer parcellation scales.}
    \label{fig:landscape_matched}
\end{figure}

The energy landscape collapsed substantially at finer parcellations even when driver-set size was matched proportionally across scales (Fig.~\ref{fig:landscape_matched}). At 68 regions, a mean of $96.17\%$ of sampled sets were within 5\% of the best observed energy. This fraction rose to $99.74\%$ at 114 regions and $>99.99\%$ at 219 regions. The collapse is therefore not an artifact of comparing driver-set sizes that represent a smaller fraction of the network at coarser scales: even when the fraction of nodes receiving input is held approximately constant, finer parcellations yield a near-flat scalar energy landscape in which driver-node identity is nearly irrelevant for average energy. This replicates the pattern reported in the main text using fixed absolute set sizes and reinforces the conclusion that geometric criteria are necessary to discriminate driver sets at finer parcellation scales.

\section{Robustness of the control-geometry advantage}
\label{sec:robustness}

The primary analysis used a stabilization margin of $c=0.1$, a regularization floor of $\epsilon=10^{-5}$, and an infinite-horizon Gramian with $k\in\{1,2,3,5\}$ driver nodes. We assessed the sensitivity of the geometry advantage (cycle-participation minus degree-strength difference in effective rank) to each of these choices individually.

\subsection{Regularization floor}
\label{subsec:robustness_epsilon}

The regularization floor $\epsilon$ applied to the Gramian eigenvalues affects how small eigenvalues---which correspond to hard-to-control directions---are handled. We repeated the effective-rank comparisons across seven values spanning $\epsilon\in\{10^{-8},10^{-7},10^{-6},10^{-5},10^{-4},10^{-3},10^{-2}\}$ at $c=0.01$ and fixed $k\in\{1,2,3,5\}$.

The geometry advantage was essentially unchanged from $\epsilon=10^{-8}$ through $\epsilon=10^{-5}$ (Table~\ref{tab:epsilon_sensitivity}). Mean effective-rank differences at the primary value $\epsilon=10^{-5}$ were $1.06$, $1.32$, and $1.73$ at 68, 114, and 219 regions respectively; the corresponding differences at $\epsilon=10^{-8}$ were $1.05$, $1.29$, and $1.67$. All differences remained highly significant ($p<10^{-11}$, Wilcoxon). Above $\epsilon=10^{-4}$, effective-rank differences grew substantially, because the floor progressively lifts eigenvalues in directions that are structurally hard to control, inflating the effective rank of the degree-informed set less than that of the topology-informed set. These large-$\epsilon$ values effectively remove the penalty for poor conditioning and should not be interpreted as increasing the geometry advantage in a meaningful sense; they reflect regularization dominating the signal. Conclusions drawn from the main analysis are therefore robust to the specific floor used within the range $\epsilon\in[10^{-8},10^{-5}]$.

\begin{table}[h]
\centering
\caption{\textbf{Effective-rank difference (cycle minus degree) as a function of regularization floor $\epsilon$.} Values are means across 70 subjects, averaged over $k\in\{1,2,3,5\}$; $c=0.01$, infinite-horizon Gramian. All entries with $\epsilon\leq10^{-5}$ were significant at $p<10^{-11}$ (Wilcoxon signed-rank, unadjusted).}
\label{tab:epsilon_sensitivity}
\begin{tabular}{lccc}
\toprule
$\epsilon$ & 68 regions & 114 regions & 219 regions \\
\midrule
$10^{-8}$ & 1.053 & 1.294 & 1.669 \\
$10^{-7}$ & 1.053 & 1.294 & 1.670 \\
$10^{-6}$ & 1.054 & 1.296 & 1.676 \\
$10^{-5}$ & 1.061 & 1.317 & 1.732 \\
$10^{-4}$ & 1.122 & 1.486 & 2.225 \\
$10^{-3}$ & 1.663 & 3.288 & 9.265 \\
$10^{-2}$ & 7.700 & 24.110 & 82.463 \\
\bottomrule
\end{tabular}
\end{table}

\subsection{Stabilization margin}
\label{subsec:robustness_c}

The stabilization margin $c$ shifts all eigenvalues of the normalized adjacency matrix by $-c$, ensuring the system is asymptotically stable. We compared two alternative values, $c=0.01$ (weak stabilization) and $c=1.0$ (strong stabilization), at $\epsilon=10^{-5}$ and $k\in\{1,2,3,5\}$.

\paragraph{$c=0.01$.} The geometry advantage was fully preserved and modestly larger than at the primary $c=0.1$. Mean effective-rank differences were $1.06$, $1.32$, and $1.73$ at 68, 114, and 219 regions respectively, with 69--70 of 70 subjects positive at each scale (all $p<10^{-11}$).

\paragraph{$c=1.0$.} The infinite-horizon geometry advantage was substantially attenuated at large stabilization and reversed at finer scales. Mean effective-rank differences were $-0.016$ (34/70 positive, $p=0.39$), $-0.052$ (24/70, $p<0.01$), and $-0.237$ (4/70, $p<10^{-11}$) at 68, 114, and 219 regions respectively. The reversal arises because at $c=1.0$ the stabilization shift is comparable in magnitude to the normalized edge weights, which lie in $[0,1]$. The shifted state matrix $A = A_0 - (\lambda_{\max}+c)I$ is dominated by the $-cI$ term, causing the system dynamics to approach uniform, isotropic decay in which all nodes are structurally interchangeable. In this regime, degree-informed and topology-informed sets are equally matched geometrically, and the small structural differences that drive the topology advantage at lower $c$ are effectively erased. When the same $c=1.0$ system is analyzed with finite-horizon Gramians (see Section~\ref{subsec:robustness_horizon}), the geometry advantage is fully restored, confirming that the reversal reflects a degenerate operating regime of the infinite-horizon Gramian under aggressive stabilization rather than a genuine failure of cycle participation to identify geometrically advantaged driver sets.

\subsection{Finite-horizon Gramians}
\label{subsec:robustness_horizon}

The main text used the infinite-horizon controllability Gramian. We repeated the driver-set comparisons using finite-horizon Gramians at $T\in\{1.0,2.5,5.0\}$ time units, evaluated at both $c=0.01$ and $c=1.0$ with $\epsilon=10^{-8}$.

The geometry advantage was preserved and substantially enlarged under all finite horizons and both stabilization values (Table~\ref{tab:horizon_sensitivity}). At $T=5$, $c=0.01$, mean effective-rank differences were $4.97$, $13.19$, and $34.26$ at 68, 114, and 219 regions respectively, with 70/70 subjects positive at every scale and horizon. At $c=1.0$, the reversal seen for the infinite-horizon Gramian was absent: mean effective-rank differences at $T=1$ were $1.09$, $2.28$, and $3.95$ (all 70/70 positive, $p<10^{-12}$). Finite-horizon Gramians are less sensitive to the aggressive stabilization because integration over a finite window retains more information about the structural differences among driver-node configurations. The consistently larger differences under finite horizons indicate that the geometric advantage of topology-informed driver sets is, if anything, understated by the infinite-horizon analysis used in the main text.

\begin{table}[h]
\centering
\caption{\textbf{Effective-rank difference (cycle minus degree) across finite horizons and stabilization values.} Values are means across 70 subjects, averaged over $k\in\{1,2,3,5\}$; $\epsilon=10^{-8}$. All entries are significant at $p<10^{-12}$ (Wilcoxon signed-rank, unadjusted), with 70/70 subjects positive in all cells.}
\label{tab:horizon_sensitivity}
\begin{tabular}{lcccccc}
\toprule
& \multicolumn{3}{c}{$c=0.01$} & \multicolumn{3}{c}{$c=1.0$} \\
\cmidrule(lr){2-4}\cmidrule(lr){5-7}
Horizon $T$ & 68 & 114 & 219 & 68 & 114 & 219 \\
\midrule
1.0 & 1.76 & 4.43 & 9.50 & 1.09 & 2.28 & 3.95 \\
2.5 & 3.73 & 9.76 & 23.16 & 1.42 & 2.92 & 5.05 \\
5.0 & 4.97 & 13.19 & 34.26 & 1.44 & 2.94 & 5.09 \\
\bottomrule
\end{tabular}
\end{table}

\subsection{Larger driver-set sizes}
\label{subsec:robustness_largek}

The main analysis used $k\in\{1,2,3,5\}$. We extended the comparison to all $k$ values on a grid that included $k=10$ and approximately 5\%, 10\%, and 20\% of each network, giving $k\in\{1,2,3,5,7,10,14\}$ at 68 regions, $\{1,2,3,5,6,10,11,23\}$ at 114 regions, and $\{1,2,3,5,10,11,22,44\}$ at 219 regions.

The effective-rank advantage of cycle-participation driver sets grew monotonically with $k$ at all three scales (Table~\ref{tab:largek_sensitivity}). At 68 regions, the advantage grew from $0.45$ at $k=1$ to $2.19$ at $k=10$; at 219 regions it grew from $0.42$ at $k=1$ to $8.24$ at $k=44$. Larger driver sets allow more driver nodes to be drawn from the cycle-participation distribution, progressively sampling more of the mesoscale loop structure identified by persistent homology. The monotone scaling is consistent with the interpretation that cycle participation identifies a complementary axis of driver-node quality: adding more topology-informed nodes continues to expand the controllable subspace into new eigen-directions rather than overlapping with directions already accessed by earlier selections.

\begin{table}[h]
\centering
\caption{\textbf{Effective-rank difference (cycle minus degree) as a function of driver-set size $k$.} Values are means across 70 subjects; $c=0.01$, $\epsilon=10^{-5}$, infinite-horizon Gramian. $n_+$ is the number of subjects with a positive difference. All entries significant at $p<10^{-9}$ (Wilcoxon signed-rank, unadjusted).}
\label{tab:largek_sensitivity}
\begin{tabular}{lcc|lcc|lcc}
\toprule
\multicolumn{3}{c|}{68 regions} & \multicolumn{3}{c|}{114 regions} & \multicolumn{3}{c}{219 regions} \\
$k$ & Mean diff. & $n_+$/70 & $k$ & Mean diff. & $n_+$/70 & $k$ & Mean diff. & $n_+$/70 \\
\midrule
1  & 0.45 & 60 & 1  & 0.40 & 60 & 1  & 0.42 & 62 \\
2  & 0.90 & 64 & 2  & 1.09 & 67 & 2  & 1.42 & 69 \\
3  & 1.30 & 69 & 3  & 1.63 & 69 & 3  & 2.01 & 70 \\
5  & 1.59 & 69 & 5  & 2.16 & 70 & 5  & 3.08 & 70 \\
7  & 1.97 & 68 & 6  & 2.40 & 70 & 10 & 5.24 & 70 \\
10 & 2.19 & 70 & 10 & 3.05 & 70 & 11 & 5.64 & 70 \\
14 & 2.14 & 69 & 11 & 2.98 & 70 & 22 & 6.96 & 70 \\
   &      &    & 23 & 3.65 & 69 & 44 & 8.24 & 70 \\
\bottomrule
\end{tabular}
\end{table}

\section{Relationship between cycle participation and betweenness centrality}
\label{sec:betweenness}

Because nodes that bridge many mesoscale loops might also lie on many shortest paths, we asked whether cycle participation is redundant with betweenness centrality. For each subject and parcellation scale we computed weighted betweenness centrality (shortest-path lengths taken as inverse edge weights) and its within-subject Spearman rank correlation with cycle participation, degree strength, and between the latter two (Table~\ref{tab:betweenness}).

Cycle participation was only weakly-to-moderately correlated with betweenness (mean Spearman $\rho = 0.31$--$0.37$ across scales), of a similar magnitude to its correlation with degree strength ($\rho = 0.39$--$0.48$). By contrast, betweenness was strongly correlated with degree strength ($\rho = 0.74$--$0.77$). Cycle participation is therefore less redundant with classical node-level centrality than betweenness is with degree: the two conventional measures largely recapitulate one another, whereas cycle participation indexes an axis of node importance---mesoscale loop structure identified by persistent homology---that is not reducible to either. This is consistent with the divergence between topology-based and degree-based rankings that motivates the main analysis.

\begin{table}[h]
\centering
\caption{\textbf{Within-subject rank correlations among cycle participation, betweenness centrality, and degree strength.} Entries are the mean Spearman $\rho$ $\pm$ standard deviation across 70 subjects, computed separately at each parcellation scale.}
\label{tab:betweenness}
\begin{tabular}{lccc}
\toprule
Measure pair & 68 regions & 114 regions & 219 regions \\
\midrule
Cycle participation vs.\ betweenness   & $0.37 \pm 0.11$ & $0.31 \pm 0.11$ & $0.35 \pm 0.07$ \\
Cycle participation vs.\ degree strength & $0.48 \pm 0.12$ & $0.39 \pm 0.12$ & $0.40 \pm 0.09$ \\
Betweenness vs.\ degree strength       & $0.77 \pm 0.05$ & $0.77 \pm 0.04$ & $0.74 \pm 0.04$ \\
\bottomrule
\end{tabular}
\end{table}

\section{Sensitivity of cycle participation to the choice of $H_1$ representative}
\label{sec:representatives}

Cycle participation is defined through the vertices a representative touches, and a
homology class admits many representatives. We therefore recomputed the score at 114
regions under three constructions: the Ripser cocycles used in the main text, homology
pairing cycles from GUDHI (birth-edge and death-triangle vertices), and length-minimized
cycles obtained by deleting the birth edge and taking the shortest replacement path at
the birth filtration scale. Driver sets were evaluated at $k\in\{2,3,5,15\}$; $k=1$ was
excluded because the two criteria often select the same node.

The two persistence computations agreed to numerical precision---same feature counts,
barcodes to $3\times10^{-8}$, and the same birth simplex for all 7,866 matched features
---so differences below reflect the representative convention rather than the software.
The constructions induced substantially different scores (median between-construction
Spearman $\rho = 0.40$--$0.68$) and driver sets (mean Jaccard $0.20$--$0.27$ at $k=15$),
but the effective-rank advantage over degree was positive and significant under all
three at every $k$ (Table~\ref{tab:representatives}), as was the participation-ratio
advantage (median differences at $k=15$: $+6.28$, $+2.01$, $+0.29$).

What the representative changes is magnitude, and it does so systematically: as
representatives become more localized, their correlation with degree rises ($\rho=0.38$,
$0.55$, $0.76$), their driver sets move toward the degree set (Jaccard $0.18$, $0.30$,
$0.50$; chamfer distance $22.7$, $14.9$, $9.8$~mm), and the advantage shrinks. Shortest
cycles preferentially route through high-degree hubs, which is why minimizing length
erodes the contrast with degree. The distributed cohomological representative used in
the main text is thus the construction that separates the two most clearly, though the
direction of the geometric effect does not depend on that choice.

\begin{table}[h]
\centering
\caption{\textbf{Effective-rank difference (cycle minus degree) under three $H_1$
representative constructions.} Median across 70 subjects at 114 regions, two-sided
Wilcoxon $p$ in parentheses; $c=0.1$, $\epsilon=10^{-5}$, infinite-horizon Gramian.}
\label{tab:representatives}
\begin{tabular}{lccc}
\toprule
$k$ & Cocycles & Pairing cycles & Geodesic cycles \\
\midrule
2  & $+0.44$ ($5\!\times\!10^{-11}$) & $+0.55$ ($2\!\times\!10^{-11}$) & $+0.02$ ($2\!\times\!10^{-3}$) \\
3  & $+1.25$ ($5\!\times\!10^{-13}$) & $+0.96$ ($5\!\times\!10^{-13}$) & $+0.16$ ($8\!\times\!10^{-6}$) \\
5  & $+2.46$ ($4\!\times\!10^{-13}$) & $+1.57$ ($4\!\times\!10^{-13}$) & $+0.33$ ($2\!\times\!10^{-8}$) \\
15 & $+5.74$ ($4\!\times\!10^{-13}$) & $+2.62$ ($4\!\times\!10^{-13}$) & $+0.53$ ($3\!\times\!10^{-8}$) \\
\bottomrule
\end{tabular}
\end{table}

\end{document}